\documentclass[11pt,a4j]{article}
\usepackage{amsmath,amssymb}
\usepackage{bm}
\usepackage{graphicx}
\usepackage{verbatim}
\usepackage{wrapfig}
\usepackage{ascmac}
\usepackage{relsize,amsmath,amssymb}
\usepackage{subcaption}
\usepackage{float}
\usepackage{mathrsfs}
\usepackage{fancybox}
\usepackage{here}
\usepackage{cite}
\newcommand{\eps}{\epsilon}

\newcommand\la{\langle}
\newcommand\ra{\rangle}

\newcommand\beq{\begin{eqnarray}}
\newcommand\eeq{\end{eqnarray}}
\newcommand\beqs{\begin{eqnarray*}}
\newcommand\eeqs{\end{eqnarray*}}

\def\wslash{\rlap/{\mkern-1mu w}}

\def\Nhat{\widehat{N}}

\def\nn{\nonumber}
\def\del{\partial}

\def\oz{\frac{1}{z}}
\def\ozd{\frac{1}{z'}}
\def\ozo{\frac{1}{z_1}}
\def\ozt{\frac{1}{z_2}}

\def\Nhat{\widehat{N}}
\def\Ohat{\widehat{O}}

\def\3Tb{3\bar{T}}

\allowdisplaybreaks[1]

\begin{document}
\begin{flushright}
\today
\end{flushright}
\vspace*{5mm}
\begin{center}
{\Large \bf A New Derivation of the Twist-3 Gluon Fragmentation\\[10pt]
Contribution to
Polarized Hyperon Production}
\vspace{1.5cm}\\
 {\sc Riku Ikarashi$^1$, Yuji Koike$^2$, Kenta Yabe$^1$ and Shinsuke Yoshida$^{3,4}$}
\\[0.7cm]
\vspace*{0.1cm}
{\it $^1$ Graduate School of Science and Technology, Niigata University,
Ikarashi, Niigata 950-2181, Japan}

\vspace{0.2cm}
{\it $^2$ Department of Physics, Niigata University, Ikarashi, Niigata 950-2181, Japan}

\vspace{0.2cm}

{\it $^3$ Guangdong Provincial Key Laboratory of Nuclear Science, Institute of Quantum Matter, South
China Normal University, Guangzhou 510006, China}

\vspace{0.2cm}

{\it $^4$ Guangdong-Hong Kong Joint Laboratory of Quantum Matter, Southern Nuclear Science Computing Center, South China Normal University, Guangzhou 510006, China}
\\[3cm]

{\large \bf Abstract} \end{center}
\vspace{0.2cm}

A novel method of formulating the twist-3 gluon fragmentation function contribution
to hyperon polarization in the proton-proton collision is presented.
The method employs a covariant gauge and takes full advantage of the Ward-Takahashi identities
before performing the collinear expansion.
It provides a robust way of constructing the general cross section formula, and
also a clear understanding for the absence of the ghost-like terms in the twist-3 cross 
section in the leading order with respect to the QCD coupling constant.

%
%
\newpage
\section{Introduction}

\begin{figure}[h]
\begin{center}
\includegraphics[width=16cm]{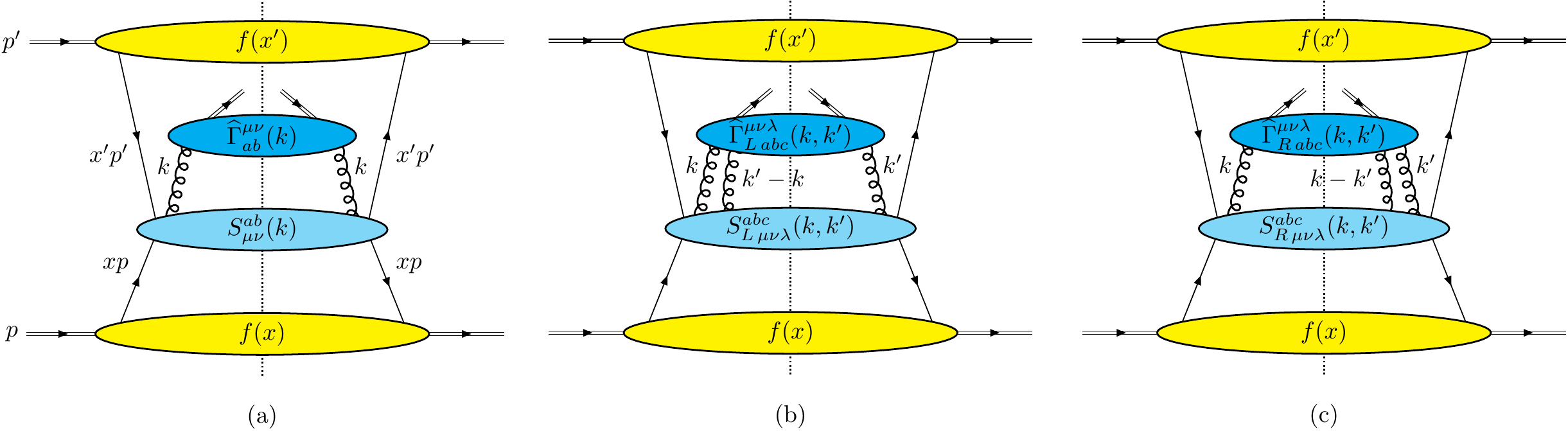}
\end{center}
\caption{Generic diagrams representing the twist-3 gluon FF contribution to 
$pp\to\Lambda^\uparrow X$ for the quark and antiquark distributions in the initial 
unpolarized protons.  The gluon distribution functions in the initial protons also contribute.  
The diagrams (a), (b) and (c), respectively,  correspond to $W^{\rm (a)}_g$, $W^{\rm (b)}_g$ 
and $W^{\rm (c)}_g$ in (\ref{s97}).  
}
\label{fig3}
\end{figure}

In our recent paper \cite{Koike:2021awj}, we presented a formalism for calculating the twist-3
gluon fragmentation function (FF) contribution to
the polarized hyperon production in the proton-proton collision:
\begin{eqnarray}
p(p)+p(p')\to\Lambda^\uparrow(P_h,S_\perp)+X,
\label{pplambdax}
\end{eqnarray}
where $p$, $p'$ and $P_h$ are the momenta of the particles and $S_\perp$ is the transverse 
spin vector of final $\Lambda^\uparrow$.  
This contribution is diagrammatically
shown in Fig. \ref{fig3}, and the corresponding
twist-3 cross section can be calculated from the following formula:
\footnote{To get a gauge- and frame-independent twist-3 cross section, the $q\bar{q}g$-type
FF contribution shown in Fig. 2 of \cite{Koike:2021awj} needs to be added to
(\ref{PPgfragForma}). 
Since the calculation of the contribution is straightforward, it is not considered in this paper.
Readers should refer to \cite{Koike:2021awj} for that contribution.  }
\begin{eqnarray}
&&\hspace{-0.5cm} E_{h}\frac{d\sigma(p,p',P_h;S_\perp)}{d^3P_h}=
\frac{1}{16\pi^2S_E}\sum_{i,j=q,\bar{q},g}\int^1_0\frac{dx}{x}f_i(x)
\int^1_0 \frac{dx'}{x'}f_j(x')\nn\\
&&\times\biggl[
{\Omega^\mu_{\ \alpha}}{\Omega^\nu_{\ \beta}}
\int^1_0 dz\, {\rm Tr}\left[\widehat{\Gamma}^{\alpha\beta}(z)S_{\mu\nu}(P_h/z)\right]
-i\,{\Omega^\mu_{\ \alpha}}{\Omega^\nu_{\ \beta}}{\Omega^\lambda_{\ \gamma}}
\int^1_0 dz\, {\rm Tr}\left[\widehat{\Gamma}_{\del}^{\alpha\beta\gamma}(z)
\left.\frac{\del S_{\mu\nu}(k)}{\del k^\lambda}\right|_{\rm c.l.}\right]\nn\\
&&+
{\Re}\Bigl\{i\,
{\Omega^\mu_{\ \alpha}}{\Omega^\nu_{\ \beta}}{\Omega^\lambda_{\ \gamma}}
\int^1_0 \frac {dz}{z}\int^\infty_z \frac{dz'}{z'}\,
\left(\frac{1}{1/z-1/z'}\right)
{\rm Tr}\left[
\widehat{\Gamma}_{F\,abc}^{\alpha\beta\gamma}\left(\ozd,\oz\right)
S^{abc}_{L\,\mu\nu\lambda}\left({1\over z'},{1\over z}\right)\right]
\Bigr\}\biggr],
\label{PPgfragForma}
\end{eqnarray}
where $f_i(x)$ ($i=q,\,\bar{q},\,g$) is the twist-2 unpolarized quark, antiquark, and gluon distributions
in the unpolarized proton with the parton's momentum fraction $x$, 
$S_E=(p+p')^2$ is the center-of-mass energy squared, and ${\rm Tr}$ indicates the sum over
all spinor or Lorentz indices depending on the channels.  
The correlation functions
$\widehat{\Gamma}^{\alpha\beta}(z)$, $\widehat{\Gamma}_\partial^{\alpha\beta\gamma}(z)$
and $\widehat{\Gamma}_{F\,abc}^{\alpha\beta\gamma}\left(\ozd,\oz\right)$, respectively, define
{\it intrinsic}, {\it kinematical} and {\it dynamical}
twist-3 gluon FFs. (For the precise definition, see section 2.)
$S_{\mu\nu}(k)$ and $S^{abc}_{L\,\mu\nu\lambda}(z',z)$ are the partonic hard parts which are,
at the beginning, 
convoluted with the Fourier transform of the hadronic matrix elements
$\sim \la 0| A_b^\nu (0)|hX\ra \la hX|A_a^\mu (\xi)|0\ra$
and
$\sim \la 0| A_b^\nu (0)|hX\ra \la hX|A_a^\mu (\xi)gA_c^\lambda (\eta )|0\ra$, respectively.  
The symbol
$\Omega^\mu_{\ \alpha}$ is defined as $\Omega^\mu_{\ \alpha}=g^\mu_{\ \alpha} -P_h^\mu w_\alpha$
with another lightlike vector $w$ satisfying $P_h\cdot w=1$, and
$\left. \right|_{\rm c.l.}$ implies the collinear limit, $k\to P_h/z$.  
In \cite{Koike:2021awj}, the formula (\ref{PPgfragForma}) was applied to the process
(\ref{pplambdax}) and the cross section was calculated in the leading order (LO) with respect to the
QCD coupling constant.  
This completed the LO twist-3 cross section 
for (\ref{pplambdax}) 
together with
the known results for the contribution from the twist-3 distribution 
function\,\cite{Kanazawa:2001a,Zhou:2008,Koike:2015zya}
and the twist-3 quark fragmentation function\,\cite{Koike:2017fxr}.
Since the formula (\ref{PPgfragForma}) is a very general one,
it can be easily adopted for other processes such as 
$e^+e^-\to \Lambda^\uparrow X$\cite{Gamberg:2018fwy} and
$ep\to e\Lambda^\uparrow X$\cite{Koike:2022ddx,Ikarashi:2022yzg}, etc.

To derive the general formula (\ref{PPgfragForma}), we applied in \cite{Koike:2021awj} the 
collinear expansion
to the hard parts $S_{\mu\nu}(k)$ and $S_{L\,\mu\nu\lambda}^{abc}(k,k')$.  
Using the Ward-Takahashi identities for the partonic hard parts, 
we could eventually rewrite the twist-3 cross sections
in terms of the low derivatives of the hard parts and the
gauge invariant correlation functions of the gluon's field strengths as in (\ref{PPgfragForma}).  
Actual calculation, however, is extremely complicated and lengthy, and is not easy to
see how the correlation functions of the gauge field $A_a^\mu$ is converted into
those of the field strength $F^a_{\mu\nu}=\partial_\mu A^a_\nu -\partial_\nu A^a_\mu
+gf^{abc}A^b_\mu A^c_\nu$.  
Furthermore, 
vanishing of the ghost-like terms appearing in the Ward-Takahashi identities
is essential to reach (\ref{PPgfragForma}), which was not clearly shown in \cite{Koike:2021awj}.  
Therefore an easier way of deriving (\ref{PPgfragForma})
is very useful.

In this paper, we present a much more robust and concise way of deriving
 (\ref{PPgfragForma}).  In this method, we use Ward-Takahashi identities from the outset to
convert gauge fields into a part of the field strengths, which results in substantial saving in 
actual calculation.  This procedure was once adopted for deriving the twist-3 three-gluon
distribution contribution to the single spin asymmetry in $ep^\uparrow\to eDX$,
where three-gluon distribution contributes as an only source for the asymmetry
and appears as a "pole contribution"\cite{Beppu:2010qn}.   
For the present case of the twist-3 gluon FF for
(\ref{pplambdax}), three types of FFs contribute as a "nonpole contribution",  and hence the situation 
is much more complicated.  
Furthermore, our present method provides a clear proof for the absence of
the ghost-like terms which appear in the Ward-Takahashi identities.  This is crucial
to guarantee the gauge invariance of the twist-3 cross section.

The remainder of the paper is organized as follows:
In section 2, a brief summary of the twist-3 gluon FFs which appear in (\ref{PPgfragForma}) is given. 
In section 3 and Appendix, we present a novel derivation of  (\ref{PPgfragForma}) and
prove the absence of the ghost-like terms in the LO twist-3 cross section.  
Section 4 is devoted to a brief summary.


\section{Gluon Fragmentation Functions}

Here we summarize the twist-3 gluon FFs in our notation
which appear in 
(\ref{PPgfragForma})\,\cite{Koike:2021awj}\footnote{See also 
\cite{Mulders:2000sh,Gamberg:2018fwy,Koike:2019zxc} 
for earlier references and more details about the gluon FFs.}.
The twist-3 {\it intrinsic} gluon FFs are defined as
\begin{eqnarray}
&&
\widehat{\Gamma}^{\alpha\beta}(z)\nonumber\\
&&={1 \over {N^2-1}}\sum_X\!\int \!\frac{d\lambda}{2\pi} 
e^{-i{\lambda \over z}}\la 0|\bigl([\infty w,0]F^{w\beta}(0)\bigr)_a| h(P_h,S_h)X\ra
\la h(P_h,S_h)X|\bigl(F^{w\alpha}(\lambda w)[\lambda w,\infty w]\bigr)_a| 0\ra \nn\\
&&
=-g_{\perp}^{\alpha\beta}\widehat{G}(z)-iM_h\eps^{P_h w\alpha\beta}(S_h\cdot w)
\Delta\widehat{G}(z)
-iM_h\eps^{P_h w S_{\perp} [\alpha}w^{\beta]}\Delta\widehat{G}_{3T}(z)
+M_h\eps^{P_h w S_{\perp}\{\alpha}w^{\beta\}}\Delta\widehat{G}_{3\bar{T}}(z),\nn\\
\label{gFraI}
\end{eqnarray}
where $N=3$ is the number of colors for a quark, 
$|h(P_h,S_h)\ra$ is the spin-1/2 hyperon state with the four momentum $P_h$  ($P_h^2=M_h^2$)
and the spin vector $S_h$ ($S_h^2=-M_h^2$), and $[\lambda w, \infty w]$ is the gauge link
in the adjoint representation connecting $\lambda w$ and $\infty w$. 
For the transversely polarized baryon,
we use the spin vector $S_\perp$ normalized as $S_\perp^2=-1$.  
In the twist-3 accuracy
$P_h$ can be regarded as lightlike.  For a baryon with large momentum, 
$P_h\simeq (|\vec{P}_h|, \vec{P}_h)$,
another lightlike vector $w$ is defined as $w=1/(2|\vec{P}_h|^2)(|\vec{P}_h|, -\vec{P}_h)$ 
which satisfies $P_h\cdot w=1$.  
$\widehat{G}(z)$ and  $\Delta \widehat{G}(z)$ are twist-2, and 
$\Delta \widehat{G}_{3T}(z)$ and $\Delta \widehat{G}_{3\bar{T}}(z)$ are twist-3.  
We also note 
$\Delta \widehat{G}_{3\bar{T}}(z)$ is naively $T$-odd, contributing to hyperon polarization.  
Each function in (\ref{gFraI}) has a support on $0<z<1$.

The {\it kinematical} FFs are defined from the transverse derivative of the
correlation functions of the field strengths: 
\begin{eqnarray}
&&\widehat{\Gamma}_{\del}^{\alpha\beta\gamma}(z)
=\frac{1}{N^2-1}\sum_X\!\int \!\!\frac{d\lambda}{2\pi}\!e^{-i{\lambda\over z}}\la 0|
\bigl([\infty w,0]
F^{w\beta}(0)\bigr)_a| h(P_h,S_\perp)X\ra\nn\\
&&\qquad\qquad\qquad\qquad\qquad
\times\la h(P_h,S_\perp)X|\bigl(F^{w\alpha}(\lambda w)
[\lambda w,\infty w]\bigr)_a| 0\ra\overleftarrow{\del}^\gamma\nn\\
&&\qquad\qquad= -i\frac{M_h}{2}g_{\perp}^{\alpha\beta}\eps^{P_h w S_\perp \gamma}\widehat{G}_T^{(1)}(z)
+\frac{M_h}{2}\eps^{P_h w\alpha\beta}S_\perp^\gamma\Delta\widehat{G}_T^{(1)}(z)\nn\\
&&\qquad\qquad\qquad\qquad
-i\frac{M_h}{8}\left(\eps^{P_h w S_\perp\{\alpha}g_{\perp}^{\beta\}\gamma}
+\eps^{P_h w\gamma\{\alpha}S_\perp^{\beta\}}\right)\Delta\widehat{H}_T^{(1)}(z),
\label{gFraK}
\end{eqnarray}
where each function is defined to be real and 
has a support on $0<z<1$.   


The {\it dynamical} FFs are defined from
the lightcone correlation functions of three field strengths:
\begin{eqnarray}
&&\widehat{\Gamma}_{F\,abc}^{\alpha\beta\gamma}(\ozo,\ozt)\nn\\ 
&&=
\sum_X\!\int \!\!\frac{d\lambda}{2\pi}\! \int \!\!\frac{d\mu}{2\pi}
 e^{-i{\lambda\over z_1}}e^{-i\mu({1\over z_2}-{1\over z_1})}\la 0|F^{w\beta}_b(0)| h(P_h,S_\perp)X
\ra\la h(P_h,S_\perp)X|F^{w\alpha}_a(\lambda w)
 gF^{w\gamma}_c(\mu w)| 0\ra
\nn\\
&&=\frac{if^{abc}}{N}
\widehat{\Gamma}_{FA}^{\alpha\beta\gamma}(\ozo,\ozt)
+d^{abc}\frac{N}{N^2-4}
\widehat{\Gamma}_{FS}^{\alpha\beta\gamma}(\ozo,\ozt),
\label{gFraD}
\end{eqnarray}
where the gauge link operators are suppressed for simplicity, 
and $f^{abc}$ and $d^{abc}$ are the anti-symmetric and symmetric 
structure constants of color SU(N).  
The dynamical FFs can be defined as the decomposition of the two correlation 
functions in (\ref{gFraD}) as
\begin{eqnarray}
&&\widehat{\Gamma}_{FA}^{\alpha\beta\gamma}(\ozo,\ozt)\nn\\ 
&&=\frac{-if_{abc}}{N^2-1}\sum_X\!\int \!\!\frac{d\lambda}{2\pi}\! \int \!\!\frac{d\mu}{2\pi}
 e^{-i{\lambda\over z_1}}e^{-i\mu({1\over z_2}-{1\over z_1})}\la 0|F^{w\beta}_b(0)| h(P_h,S_\perp)X
\ra\la h(P_h,S_\perp)X|F^{w\alpha}_a(\lambda w)
 gF^{w\gamma}_c(\mu w)| 0\ra\nn\\ 
&&=-{M_h}\biggl(\Nhat_{1}\left(\ozo,\ozt\right)g^{\alpha\gamma}_{\perp}
\eps^{P_h w S_\perp\beta}\hspace{-0.05cm}+\Nhat_{2}\left(\ozo,\ozt\right)g^{\beta\gamma}_{\perp}
\eps^{P_h w S_\perp\alpha}\hspace{-0.05cm}-\hspace{-0.05cm}
\Nhat_{2}\left(\ozt-\ozo,\ozt\right)g^{\alpha\beta}_{\perp}\eps^{P_h w S_\perp\gamma} \biggr), 
\label{gFraDA}\nn\\
\end{eqnarray}
\begin{eqnarray}
&&\widehat{\Gamma}_{FS}^{\alpha\beta\gamma}(\ozo,\ozt)\nn\\ 
&&=\frac{d_{abc}}{N^2-1}\sum_X
\!\int \!\!\frac{d\lambda}{2\pi}\! \int \!\!\frac{d\mu}{2\pi} e^{-i{\lambda\over z_1}}
e^{-i\mu({1\over z_2}-{1\over z_1})}\la 0|F^{w\beta}_b(0)| h(P_h,S_\perp)
X\ra\la h(P_h,S_\perp)X|F^{w\alpha}_a(\lambda w)gF^{w\gamma}_c(\mu w)| 0\ra\nn\\
&&=-{M_h}\biggl(\Ohat_{1}\left(\ozo,\ozt\right)g^{\alpha\gamma}_{\perp}
\eps^{P_h w S_\perp\beta}\hspace{-0.05cm}+\Ohat_{2}\left(\ozo,\ozt\right)g^{\beta\gamma}_{\perp}
\eps^{P_h w S_\perp\alpha}\hspace{-0.05cm}+
\Ohat_{2}\left(\ozt-\ozo,\ozt\right)g^{\alpha\beta}_{\perp}\eps^{P_h w S_\perp\gamma} \biggr).\nn\\
\label{gFraDS}
\end{eqnarray}
Correlation functions (\ref{gFraDA}) and (\ref{gFraDS}), respectively, define
two independent set of the {\it complex} functions 
$\left\{\Nhat_{1},
\Nhat_{2}
\right\}$
and $\left\{\Ohat_{1}
, \Ohat_{2}
\right\}$ due to the exchange symmetry of the field strengths.  
Functions $\Nhat_1$ and $\Ohat_1$
satisfy the relations
\beq
\Nhat_{1}\left(\ozo,\ozt\right)&=&-\Nhat_{1}\left(\ozt-\ozo,\ozt\right),\nn\\
\Ohat_{1}\left(\ozo,\ozt\right)&=&\Ohat_{1}\left(\ozt-\ozo,\ozt\right).  
\label{swi_nat}
\eeq
The real parts of these four FFs are $T$-even and the imaginary parts are $T$-odd, 
the latter being the sources of SSAs.  
$\Nhat_{1,2}\left(\ozo,\ozt\right)$ and $\Ohat_{1,2}\left(\ozo,\ozt\right)$
have a support on $\ozt>1$ and $\ozt>\ozo>0$.

\section{Twist-3 gluon fragmentation contribution to $pp\to\Lambda^\uparrow X$}

In this section we present a robust way to derive the basic formula (\ref{PPgfragForma}).  
The twist-3 gluon FF contribution to (\ref{pplambdax}) can be written as
\begin{eqnarray}
E_{h}\frac{d\sigma(p,p',P_h;S_\perp)}{d^3P_h}=\frac{1}{16\pi^2S_{E}}
\sum_{i,j=q,\bar{q},g}
\!\int\!\!
\frac{dx}{x}f_i(x)\!\int\frac{dx'}{x'}f_j(x')
W_{g}(xp,x'p',P_h/z,S_\perp),
\label{s21}
\end{eqnarray}
where 
$W_g$ represents the partonic hard scattering
followed by the fragmentation of a gluon into the final $\Lambda^\uparrow$. 
Figure \ref{fig3} shows the generic structure of the LO diagrams for this contribution.  
Corresponding to 
Figs. \ref{fig3} (a), (b), (c), $W_g$ consists of three parts:
\begin{eqnarray}
\hspace{-1cm}&&W_{g}(xp,x'p',P_h;S_\perp) \equiv  W^{\rm (a)}_g+W^{\rm (b)}_g+W^{\rm (c)}_g\nn\\
&&\qquad=
\int\frac{d^4k}{(2\pi)^4}
\left[
S^{ab}_{\mu\nu}(k)\widehat{\Gamma}^{\mu\nu}_{ab}(k)
\right]\nn\\
&&\qquad+
{1\over 2} \int\frac{d^4k}{(2\pi)^4}\int\frac{d^4k'}{(2\pi)^4}
\biggl[
S^{abc}_{L\,\mu\nu\lambda}(k,k')\widehat{\Gamma}^{\mu\nu\lambda}_{L\,abc}(k,k')
+ S^{abc}_{R\,\mu\nu\lambda}(k,k')\widehat{\Gamma}^{\mu\nu\lambda}_{R\,abc}(k,k')
\biggr],
\label{s97}
\end{eqnarray}
where $\widehat{\Gamma}^{\mu\nu}_{ab}(k)$, 
$\widehat{\Gamma}^{\mu\nu\lambda}_{L\,abc}(k,k')$ and 
$\widehat{\Gamma}^{\mu\nu\lambda}_{R\,abc}(k,k')$ 
are the hadronic matrix elements of the gauge (gluon) fields 
with $k$ and $k'$ the four momenta of the gluons fragmenting into the final $\Lambda$, 
and $S^{ab}_{\mu\nu}(k)$, $S^{abc}_{L\,\mu\nu\lambda}(k,k')$ and 
$S^{abc}_{R\,\mu\nu\lambda}(k,k')$
are the corresponding partonic hard scattering parts with the color indices
$a,\,b,\,c$ and the Lorentz indices $\mu,\,\nu,\lambda$.  
In (\ref{s97}), the factor $1/2$ in front of
$W_g^{\rm (b)}$ and $W_g^{\rm (c)}$ takes into account
the exchange symmetry of the gluon fields in the fragmentation matrix elements.  
Hadronic matrix elements are defined as 
\begin{eqnarray}
&&\widehat{\Gamma}^{\mu\nu}_{ab}(k)=\sum_X\int{d^4\xi}\,{e^{-ik\xi}}
{\la0|}A^\nu_b(0){|hX 
\ra}{\la hX|}A^\mu_a(\xi){|0\ra},
\\[0.4cm]
&&\widehat{\Gamma}^{\mu\nu\lambda}_{L\,abc}(k,k')
=\sum_X\int{d^4\xi}\int{d^4\eta}\,{e^{-ik\xi}}{e^{-i(k'-k)\eta}}
{\la0|}A^\nu_b(0){|hX 
\ra}{\la hX|}A^\mu_a(\xi)gA^\lambda_c(\eta){|0\ra},
\\[0.4cm]
&&\widehat{\Gamma}^{\mu\nu\lambda}_{R\,abc}(k,k')
=\sum_X\int{d^4\xi}\int{d^4\eta}\,{e^{-ik\xi}}{e^{-i(k'-k)\eta}}
{\la0|}A^\nu_b(0)gA^\lambda_c(\eta){|hX 
\ra}{\la hX|}A^\mu_a(\xi){|0\ra},
\end{eqnarray}
where the gauge coupling $g$ associated with the attachment of 
the extra gluon line to the hard part
is included in 
$\widehat{\Gamma}_{L\,abc}^{\mu\nu\lambda}(k,k')$ and 
$\widehat{\Gamma}_{R\,abc}^{\mu\nu\lambda}(k,k')$.  Therefore
the hard parts $S_{\mu\nu}^{ab}(k)$, $S_{L\,\mu\nu\lambda}^{abc}(k,k')$ and 
$S_{R\,\mu\nu\lambda}^{abc}(k,k')$ 
are of $O(g^4)$ in the LO
calculation.    
From hermiticity, one has $\widehat{\Gamma}^{\mu\nu\lambda}_{L\,abc}(k,k') ^\star
=\widehat{\Gamma}^{\nu\mu\lambda}_{R\,bac}(k',k)$ and
$S^{abc}_{L\,\mu\nu\lambda}(k,k')^\star = S^{bac}_{R\,\nu\mu\lambda}(k',k)$,
which guarantees the reality of $W_g$.  
A standard procedure to extract the twist-3 effect is the collinear expansion
of the hard parts with respect to $k$ and $k'$ around $P_h$.  We followed the method
in \cite{Koike:2021awj}
to get (\ref{PPgfragForma}).

Here we present an alternative method which leads to (\ref{PPgfragForma}) more easily.  
In this method we fully use the Ward-Takahashi identities for the hard parts
to convert some of the gluon field $A^a_\mu$ into a part of the field strength $F^a_{\mu\nu}$.
Ward-Takahashi identities for the hard part read
\begin{eqnarray}
{k}^{\mu}S^{ab}_{\mu\nu}(k)={k}^{\nu}S^{ab}_{\mu\nu}(k)=0, 
\label{ward1}
\end{eqnarray}
\begin{eqnarray}
{(k'-k)}^{\lambda}S^{abc}_{L\,\mu\nu\lambda}(k,k')&=&\frac{-if^{abc}}{N^2-1}S_{\mu\nu}(k')
+G^{abc}_{\mu\nu}(k,k'),
\label{ward2}\\
{k}^{\mu}S^{abc}_{L\,\mu\nu\lambda}(k,k')&=&\frac{if^{abc}}{N^2-1}S_{\lambda\nu}(k')
+G^{cba}_{\lambda\nu}(k'-k,k'), 
\label{ward3}\\
{k'}^{\nu}S^{abc}_{L\,\mu\nu\lambda}(k,k')&=&0, 
\label{ward4}
\end{eqnarray}
\begin{eqnarray}
{(k'-k)}^{\lambda}S^{abc}_{R\,\mu\nu\lambda}(k,k')&=&\frac{if^{abc}}{N^2-1}
S_{\mu\nu}(k)-\left(G^{bac}_{\nu\mu}(k',k)\right)^\star,
\label{ward5}\\
{k'}^{\nu}S^{abc}_{R\,\mu\nu\lambda}(k,k')&=&\frac{if^{abc}}{N^2-1}S_{\mu\lambda}(k)+
\left(G^{cab}_{\lambda\mu}(k-k',k)\right)^\star,
\label{ward6}\\
{k}^{\mu}S^{abc}_{R\,\mu\nu\lambda}(k,k')&=&0, 
\label{ward7}
\end{eqnarray}
where
$S_{\mu\nu}(k)\equiv S_{\mu\nu}^{ab}(k)\delta_{ab}$.   The $G$-terms
are the ghost-like terms which appear 
due to the off-shellness and the nonphysical polarization
of the gluon lines entering the fragmentation matrix elements.
Actual forms of those ghost-like terms for $pp\to\Lambda^\uparrow X$
were given in Appendix A of \cite{Koike:2021awj} in the LO with respect to the QCD coupling. 
They are proportional to $f^{abc}$ and satisfy the relation
\beq
k^\mu G_{\mu\nu}^{abc}(k,k')=k'^\nu G_{\mu\nu}^{abc}(k,k')=0.
\label{ghostward}
\eeq
We will see that use of the relations (\ref{ward1})-(\ref{ghostward}) from 
the outset brings enormous saving in the
actual calculation and clearer understanding on the absence of the ghost-like terms
in the LO twist-3 cross 
section\footnote{Absence of the ghost term contribution to the twist-3 cross sections
was discussed for the 3-gluon distribution contribution to
$\vec{p}p^\uparrow\to DX$\,\cite{Hatta:2013wsa} 
and twist-3 quark FF contribution to $ep^\uparrow \to e\pi X$\,\cite{Kanazawa:2013uia}.}.

We first consider $W_g^{\rm (a)}$.  The integration momentum $k$ can be decomposed as 
\beq
k^\mu=(k\cdot w)P_h^\mu+\Omega^\mu_{\ \nu}k^\nu,
\label{kdecomp}
\eeq
where $\Omega^\mu_{\ \nu}\equiv g^\mu_{\ \nu} -P_h^\mu w_\nu$.
Inserting (\ref{kdecomp}) into (\ref{ward1}), one gets
\beq
&&S_{P_h\nu}^{ab}(k) ={-1\over k\cdot w} \Omega^\mu_{\ \kappa}k^\kappa S_{\mu\nu}^{ab}(k),
\label{ward001}\\
&&S_{\mu P_h}^{ab}(k) ={-1\over k\cdot w} \Omega^\nu_{\ \tau} k^\tau S_{\mu\nu}^{ab}(k).
\label{ward002}
\eeq
Then we can write
\beq
&&S_{\mu\nu}^{ab}(k)\widehat{\Gamma}_{ab}^{\mu\nu}(k)=
S_{\mu\nu}^{ab}(k)g^\mu_{\ \alpha}g^\nu_{\ \beta}\widehat{\Gamma}_{ab}^{\alpha\beta}(k)=
S_{\mu\nu}^{ab}(k) (P_h^\mu w_{\alpha} +\Omega^\mu_{\ \alpha})
(P_h^\nu w_{\beta} +\Omega^\nu_{\ \beta})
\widehat{\Gamma}_{ab}^{\alpha\beta}(k)\nn\\
&&={1\over (k\cdot w)^2} S_{\mu\nu}^{ab}(k)
\Omega^\mu_{\ \kappa}
\Omega^\nu_{\ \tau} (-k^\kappa w_\alpha +k\cdot w g^\kappa_{\ \alpha})
(-k^\tau w_\beta +k\cdot w g^\tau_{\ \beta}) \widehat{\Gamma}^{\alpha\beta}_{ab}(k),
\label{SGa}
\eeq
where we have used (\ref{ward001}) and (\ref{ward002}) in the last equality.
Noting that one can write
\beq
&&(-k^\kappa w_\alpha +k\cdot w g^\kappa_{\ \alpha})
(-k^\tau w_\beta +k\cdot w g^\tau_{\ \beta}) \widehat{\Gamma}^{\alpha\beta}_{ab}(k)\nn\\
&&\qquad\qquad=\sum_X\int d^4\xi e^{-ik\xi}
\la 0| F^{(0)\,\tau w}_b(0) |hX\ra \la hX|F_a^{(0)\,\kappa w}(\xi) |0\ra\equiv 
\widehat{\Gamma}_{F\,ab}^{\kappa\tau}(k),
\label{AtoF}
\eeq
where 
$F^{(0)\,\kappa\sigma}_a \equiv \partial^\kappa A_a^\sigma -\partial^\sigma A_a^\kappa$
is the $O(A)$ piece of the gluon's field strength,
one obtains
\beq
\int {d^4k\over (2\pi)^4}\left[
S_{\mu\nu}^{ab}(k)\widehat{\Gamma}_{ab}^{\mu\nu}(k)\right]
=\int {d^4k\over (2\pi)^4}\left[
{1\over (k\cdot w)^2}S_{\mu\nu}^{ab}(k)
\Omega^\mu_{\ \kappa}
\Omega^\nu_{\ \tau}
\widehat{\Gamma}_{F\,ab}^{\kappa\tau}(k)\right].
\label{2body}
\eeq
To extract the twist-3 contribution, we apply the collinear expansion to the RHS of
(\ref{2body}).
Writing $k\cdot w =1/z$, the contribution up to twist-3 from (\ref{2body}) can be
obtained as
\beq
&&W_g^{\rm (a)}=\int {d^4k\over (2\pi)^4}\left[
S_{\mu\nu}^{ab}(k)\widehat{\Gamma}_{ab}^{\mu\nu}(k)\right]^{\rm twist-3}\nn\\
&&\qquad\qquad =
\int d\left( {1\over z}\right)
z^2\,\Omega^\mu_{\ \kappa}
\Omega^\nu_{\ \tau}\left[ S_{\mu\nu}\left(z\right)
\widehat{\Gamma}_{F}^{\kappa\tau}(z)
+\Omega^\lambda_{\ \rho}\left. {\partial S_{\mu\nu}(k)\over \partial k^\lambda}\right|_{k=P_h/z}
\widehat{\Gamma}_{\partial F}^{\kappa\tau\rho}(z)\right], 
\label{(a)twist3}
\eeq
where $S_{\mu\nu}(z)\equiv S_{\mu\nu}^{ab}(P_h/z)\delta^{ab}$ and
\beq
&&\widehat{\Gamma}_{F}^{\kappa\tau}(z)= {1\over N^2-1}
\sum_X\int {d\lambda \over 2\pi}\, e^{-i\lambda/z}
\la 0| F^{(0)\,\tau w}_a(0) |hX\ra \la hX|F_a^{(0)\,\kappa w}(\lambda w) |0\ra,
\label{int0th}\\
&&\widehat{\Gamma}_{\partial F}^{\kappa\tau\rho}(z)= {1\over N^2-1}
\sum_X\int {d\lambda \over 2\pi}\, e^{-i\lambda/z}
\la 0| F^{(0)\,\tau w}_a(0) |hX\ra \la hX|(-i)\partial^\rho F_a^{(0)\,\kappa w}(\lambda w) |0\ra.
\label{kin0th}
\eeq
Equations
(\ref{int0th}) and (\ref{kin0th}) are, respectively, identified as the $O(g^0)$-parts of
(\ref{gFraI}) and (\ref{gFraK}), and (\ref{(a)twist3}) represents the lowest order contribution
to the first and second terms in (\ref{PPgfragForma}).

Next we proceed to analyze $W_g^{\rm (b)}$.  Using (\ref{kdecomp}) in (\ref{ward3}), we have
\beq
(k\cdot w)S_{L\,P_h\nu\lambda}^{abc} (k,k')+\Omega^\mu_{\ \rho}k^\rho 
S_{L\,\mu\nu\lambda}^{abc} (k,k')=
{if^{abc}\over N^2-1}S_{\lambda \nu}(k') + G_{\lambda\nu}^{cba}(k'-k,k'),
\eeq
from which we obtain
\beq
S_{L\,P_h\nu\lambda}^{abc} (k,k')
={1\over k\cdot w}\left( -\Omega^\mu_{\ \rho}k^\rho 
S_{L\,\mu\nu\lambda}^{abc} (k,k')+{if^{abc}\over N^2-1} S_{\lambda \nu}(k') + 
G_{\lambda\nu}^{cba}(k'-k,k')\right).  
\label{ward01}
\eeq
Likewise, from  (\ref{kdecomp}) and (\ref{ward4}), we have
\beq
S_{L\,\mu P_h \lambda}^{abc} (k,k')
={-1\over k'\cdot w} \Omega^\nu_{\ \tau}k'^\tau
S_{L\,\mu\nu\lambda}^{abc} (k,k').  
\label{ward02}
\eeq
As in (\ref{SGa}),  (\ref{ward01}) and (\ref{ward02}) can be used to rewrite integrand of $W_g^{\rm (b)}$ as
\beq
&&S_{L\,\mu\nu\lambda}^{abc}(k,k')\widehat{\Gamma}_{L\, abc}^{\mu\nu\lambda}(k,k')
=S_{L\,\mu\nu\lambda}^{abc}(k,k')g^\mu_{\ \kappa} g^\nu_{\ \tau} g^\lambda_{\ \sigma}
\widehat{\Gamma}_{L\, abc}^{\kappa\tau\sigma}(k,k')\nn\\
&&
=S_{L\,\mu\nu\lambda}^{abc}(k,k') \bigl(P_h^\mu w_\kappa  + \Omega^\mu_{\ \kappa}\bigr)
 \bigl(P_h^\nu w_\tau  + \Omega^\nu_{\ \tau}\bigr) 
 \bigl(P_h^\lambda w_\sigma + \Omega^\lambda_{\ \sigma}\bigr)
\widehat{\Gamma}_{L\, abc}^{\kappa\tau\sigma}(k,k') \nn\\
&&
={1\over k\cdot w}{1\over k'\cdot w}S_{L\, \mu\nu\lambda}^{abc}(k,k')
\Omega^\mu_{\ \alpha}\Omega^\nu_{\ \beta}
\bigl(-k^\alpha w_\kappa+k\cdot w g^\alpha_{\ \kappa}\bigr)
\bigl(-k'^\beta w_\tau+k'\cdot w g^\beta_{\ \tau}\bigr)\nn\\
&&\qquad\qquad\qquad\qquad\qquad\times
\bigl(P_h^\lambda w_\sigma + \Omega^\lambda_{\ \sigma}\bigr)
\widehat{\Gamma}^{\kappa\tau\sigma}_{L\,abc}(k,k')\nn\\
&&\qquad\qquad+{1\over k\cdot w}
\left( {if^{abc}\over N^2-1} S_{\sigma\tau}(k')  + G_{\sigma\tau}^{cba}(k'-k,k')\right)
\widehat{\Gamma}^{w\tau\sigma}_{L\,abc}(k,k'). 
\label{SG1}
\eeq
Similarly to (\ref{AtoF}), we have 
\beq
&&\bigl(-k^\alpha w_\kappa+k\cdot w g^\alpha_{\ \kappa}\bigr)
\bigl(-k'^\beta w_\tau+k'\cdot w g^\beta_{\ \tau}\bigr)
\widehat{\Gamma}^{\kappa\tau\sigma}_{L\,abc}(k,k')\nn\\
&&=\sum_X\int d^4\xi\int d^4\eta e^{-ik\xi}e^{-i(k'-k)\eta}
\la 0| F^{(0)\,\beta w}_b(0)|hX\ra \la hX| F^{(0)\,\alpha w}_a(\xi) gA^\sigma_c (\eta) |0\ra\nn\\
&&\equiv \widehat{\Gamma}^{\alpha\beta \sigma}_{LFA\,abc}(k,k').
\label{FLderiv}
\eeq
Using this equation, (\ref{SG1}) can be rewritten as
\beq
&&S_{L\,\mu\nu\lambda}^{abc}(k,k')\widehat{\Gamma}_{L\, abc}^{\mu\nu\lambda}(k,k')\nn\\
&&\qquad=S_{L\,\mu\nu\lambda}^{abc}(k, k') {1\over k\cdot w}{1\over k'\cdot w}
\Omega^\mu_{\ \alpha}\Omega^\nu_{\ \beta}
\left( P_h^\lambda \widehat{\Gamma}_{LFA\,abc}^{\alpha\beta w}(k,k')
+\Omega^\lambda_{\ \sigma}\widehat{\Gamma}_{LFA\,abc}^{\alpha\beta\sigma}(k,k')\right)\nn\\
&&\qquad\qquad+{1\over k\cdot w}
\left( {if^{abc}\over N^2-1}S_{\sigma\tau}(k')  + G_{\sigma\tau}^{cba}(k'-k,k')\right)
\widehat{\Gamma}^{w\tau\sigma}_{L\,abc}(k,k'). 
\label{SG3body}
\eeq
Since (\ref{SG3body}) is integrated over $k$ and $k'$ in (\ref{s97}), 
one can change the integration variable as $k \to k'-k$ in
the last term of (\ref{SG3body}) containing the ghost-like term.
Furthermore, because of
$\widehat{\Gamma}^{w\tau\sigma}_{L\,abc}(k'-k,k') = \widehat{\Gamma}^{\sigma\tau w}_{L\,cba}(k,k')$,  
one can change this term as
\beq
&&{1\over k\cdot w}G_{\sigma\tau}^{cba}(k'-k,k')\widehat{\Gamma}^{w\tau\sigma}_{L\,abc}(k,k')\nn\\
&&\qquad\longrightarrow
{1\over k'\cdot w-k\cdot w}G_{\sigma\tau}^{cba}(k,k')\widehat{\Gamma}^{\sigma\tau w}_{L\,cba}(k,k')
=
{1\over k'\cdot w-k\cdot w}G_{\sigma\tau}^{abc}(k,k')\widehat{\Gamma}^{\sigma\tau w}_{L\,abc}(k,k'). 
\eeq
Using this form for the last term in (\ref{SG3body}) and
applying the collinear expansion to the first term in (\ref{SG3body}) up to twist-3, one obtains
\beq
&&\left[S_{L\,\mu\nu\lambda}^{abc}(k,k')
\widehat{\Gamma}_{L\, abc}^{\mu\nu\lambda}(k,k')\right]^{\rm twist-3}\nn\\
&&
\qquad=z\,z'\,
\Omega^\mu_{\ \alpha}\Omega^\nu_{\ \beta}\left\{
S_{L\,\mu\nu P_h}^{abc}\left({1\over z},{1\over z'}\right)+
\Omega^\lambda_{\ \gamma}k^\gamma
\left. {\partial S_{L\, \mu\nu P_h}^{abc}(k,k')\over \partial k^\lambda}\right|_{\rm c.l.}\right.\nn\\
&&\left.\qquad\qquad\qquad\qquad\qquad\qquad\qquad\qquad
+\Omega^\lambda_{\ \gamma}k'^\gamma
\left.{\partial S_{L\, \mu\nu P_h}^{abc}(k,k')\over \partial k'^\lambda}\right|_{\rm c.l.}
\right\} 
\widehat{\Gamma}_{LFA\,abc}^{\alpha\beta w}(k,k')\nn\\
&&\qquad+z\,z'\,
\Omega^\mu_{\ \alpha}\Omega^\nu_{\ \beta}\Omega^\lambda_{\ \sigma}
S_{L\,\mu\nu\lambda}^{abc}\left({1\over z},{1\over z'}\right)
\widehat{\Gamma}_{LFA\,abc}^{\alpha\beta\sigma}(k,k')\nn\\
&&\qquad+z\, {if^{abc}\over N^2-1}S_{\sigma\tau}(k') 
\widehat{\Gamma}^{w\tau\sigma}_{L\,abc}(k,k')
+{1\over 1/z'-1/z}G_{\sigma\tau}^{abc}(k,k') \widehat{\Gamma}^{\sigma\tau w}_{L\,abc}(k,k'), 
\label{SG3body2}
\eeq
where we have set
$k\cdot w={1\over z}$ and $k'\cdot w={1\over z'}$.  
The first term of (\ref{SG3body2}) can be further rewritten by the Ward-Takahashi identity (\ref{ward2}).  
Collinear limit of (\ref{ward2}) gives
\beq
S_{L\,\mu\nu P_h}^{abc}\left({1\over z},{1\over z'}\right)
={1\over 1/z'- 1/z}\left\{ {-if^{abc}\over N^2-1} S_{\mu\nu}\left({z'}\right) 
+G_{\mu\nu}^{abc}\left({1\over z},{1\over z'}\right)
\right\},
\label{SG2}
\eeq
Similarly from the collinear limit of the first derivatives of (\ref{ward2}) with 
respect to $k$ and $k'$, one obtains
\beq
&&\hspace{-0.7cm}
\left. {\partial S_{L\, \mu\nu P_h}^{abc}(k,k')\over \partial k^\lambda}\right|_{\rm c.l.}
={1\over 1/z'-1/z}\left\{
S_{L\,\mu\nu\lambda}^{abc}\left({1\over z},{1\over z'}\right)
+
\left. {\partial G_{\mu\nu}^{abc}(k,k')\over \partial k^\lambda}\right|_{\rm c.l.}\right\},
\label{SG3}\\
&&\hspace{-0.7cm}
\left. {\partial S_{L\, \mu\nu P_h}^{abc}(k,k')\over \partial k'^\lambda}\right|_{\rm c.l.}
={1\over 1/z'-1/z}\left\{
-S_{L\,\mu\nu\lambda}^{abc}\left({1\over z},{1\over z'}\right)
+\left. {\partial G_{\mu\nu}^{abc}(k,k')\over \partial k'^\lambda}\right|_{\rm c.l.}
- {if^{abc}\over N^2-1} \left. {\partial S_{\mu\nu}(k')\over \partial k'^\lambda}\right|_{\rm c.l.}
\right\}.\nn\\
\label{SG4}
\eeq
In the last term in the RHS of (\ref{SG3body2}), 
using the relation (\ref{ghostward}) and 
following a similar procedure as (\ref{SG1}) and (\ref{FLderiv}), 
one can rewrite 
\beq
G_{\mu\nu}^{abc}(k,k') \widehat{\Gamma}^{\mu\nu w}_{L\,abc}(k,k')
=z\,z'\, G_{\mu\nu}^{abc}(k,k')\Omega^\mu_{\ \alpha}\Omega^\nu_{\ \beta}
\widehat{\Gamma}^{\alpha\beta w}_{LFA\,abc}(k,k').  
\eeq
Using this form, 
one sees the collinear expansion of the last term in (\ref{SG3body2}) yields the identical terms
as the ghost-like terms in (\ref{SG2}), (\ref{SG3}) and (\ref{SG4}).
This way one obtains the ghost-like terms (i.e., terms containg $G_{\mu\nu}^{abc}$)
in (\ref{SG3body2}) as
\beq
&&
\left[S_{L\,\mu\nu\lambda}^{abc}(k,k')
\widehat{\Gamma}_{L\, abc}^{\mu\nu\lambda}(k,k')\right]^{\rm ghost}
=2\Omega^\mu_{\ \alpha}\Omega^\nu_{\ \beta}
{zz'\over 1/z'-1/z} \left\{
G_{\mu\nu}^{abc}\left( {1\over z},{1\over z'}\right)\right.\nn\\
&&\left.\qquad\qquad\qquad\qquad
+ \Omega^\lambda_{\ \gamma}\left(
\left. k^\gamma{\partial G_{\mu\nu}^{abc}(k,k')\over \partial k^\lambda}\right|_{\rm c.l.}
+ \left. k'^\gamma{\partial G_{\mu\nu}^{abc}(k,k')\over \partial k'^\lambda}\right|_{\rm c.l.}\right)
\right\}
\widehat{\Gamma}_{LFA\,abc}^{\alpha\beta w}(k,k').
\label{ghostfinal}
\eeq
Using the actual forms of the ghost-like terms given in Appendix A of \cite{Koike:2021awj}, 
we will show in Appendix that (\ref{ghostfinal}) does not contribute to the LO twist-3 cross section.  
Hence we will discard (\ref{ghostfinal}) below.

Remaining terms in (\ref{SG3body2}) can be written as
\beq
&&\hspace{-0.3cm}
\left[S_{L\,\mu\nu\lambda}^{abc}(k,k')\widehat{\Gamma}_{L\, abc}^{\mu\nu\lambda}(k,k')\right]^{\rm twist-3}
\nn\\
&&=
\Omega^\mu_{\ \alpha}\Omega^\nu_{\ \beta}{1\over 1/z+ i\epsilon}
{z'\over 1/z'-1/z+i\epsilon}
\left( {-if^{abc}\over N^2-1} \right) S_{\mu\nu}\left({ z'}\right)
\widehat{\Gamma}_{LFA\,abc}^{\alpha\beta w}(k,k')\nn\\
&&+\Omega^\mu_{\ \alpha}\Omega^\nu_{\ \beta}\Omega^\lambda_{\ \gamma}
{1\over {1/ z}+i\epsilon}
{z'\over {1/z'}-{1/ z}+i\epsilon}
S_{L\,\mu\nu\lambda}^{abc}\left({1\over z},{1\over z'}\right)\nn\\
&&\qquad\qquad\qquad\times\left\{
(k-k')^\gamma\widehat{\Gamma}_{LFA\,abc}^{\alpha\beta w}(k,k')
+\left({1\over z'}-{1\over z}\right)\widehat{\Gamma}_{LFA\,abc}^{\alpha\beta\gamma}(k,k')
\right\}\nn\\
&&+\Omega^\mu_{\ \alpha}\Omega^\nu_{\ \beta}\Omega^\lambda_{\ \gamma}
{1\over 1/z+i\epsilon}
{z'\over 1/z'-1/z+i\epsilon}\left({-if^{abc}\over N^2-1}\right)k'^\gamma
\left. {\partial S_{\mu\nu}(k')\over \partial k'^\lambda}\right|_{\rm c.l.}
\widehat{\Gamma}_{LFA\,abc}^{\alpha\beta w}(k,k')\nn\\
&&+\Omega^\mu_{\ \alpha}\Omega^\nu_{\ \beta}
{z'^2\over 1/z +i\epsilon}\left\{ S_{\mu\nu}(z') +\Omega^\lambda_{\ \gamma}k'^\gamma
\left.
{\partial S_{\mu\nu}(k')\over \partial k'^\lambda}\right|_{\rm c.l.}\right\}\nn\\
&&\qquad\qquad\qquad\times
{if^{abc}\over N^2-1}
\left(-k'^\alpha w_\sigma+k'\cdot w g^\alpha_{\ \sigma}\right)
\left(-k'^\beta w_\tau+k'\cdot w g^\beta_{\ \tau}\right)
\widehat{\Gamma}_{L\,abc}^{w\tau\sigma}(k,k'),
\label{SG5}
\eeq
where the third term in (\ref{SG3body2}) was rewritten 
as the last term by using (\ref{ward1}) and subsequest collinear expansion of $S_{\mu\nu}(k')$.  
We have introduced
$i\epsilon$ in the denominators, which gives rise to the
future pointing  gauge links.  
Integration of (\ref{SG5}) over $k$ and $k'$ proceeds as follows.
We first write for the last term of (\ref{SG5})
\beq
&&\hspace{-0.5cm}\bigl(-k'^\alpha w_\sigma+k'\cdot w g^\alpha_{\ \sigma}\bigr)
\bigl(-k'^\beta w_\tau+k'\cdot w g^\beta_{\ \tau}\bigr)
\widehat{\Gamma}_{L\,abc}^{w\tau\sigma}(k,k')\nn\\
&&=\int d^4\xi\int d^4\eta e^{-ik \xi}e^{-i(k'-k)\eta}\sum_X
\la 0| F_b^{(0)\,w\beta}(0)|hX\ra\nn\\
&&
\times\la hX| g(\partial^wA_a^w(\xi) ) A_c^\alpha (\eta) -g(\partial^w A_a^\alpha (\xi) )A_c^w(\eta)
+ gA_a^w(\xi) F_c^{(0)\,w\alpha}(\eta) +gF_a^{(0)\,w\alpha} (\xi) A_c^w(\eta) |0\ra\qquad\nn\\
&&\equiv \widehat{\Gamma}_{LF\partial\,abc}^{\alpha\beta}(k,k').  
\eeq
Then the contribution from the first term in $\left\{\quad\right\}$ 
of the last term of (\ref{SG5}) reads
\beq
&&\hspace{-0.5cm}
\int{d^4k\over (2\pi)^4}\int{d^4k'\over (2\pi)^4}
\Omega^\mu_{\ \alpha}\Omega^\nu_{\ \beta}
{z'^2\over 1/z +i\epsilon} {if^{abc}\over N^2-1}
 S_{\mu\nu}(z') \widehat{\Gamma}_{LF\partial\,abc}^{\alpha\beta}(k,k')\nn\\
&&=\int d\left({1\over z'}\right)\int d\left({1\over z}\right) \Omega^\mu_{\ \alpha}
\Omega^\nu_{\ \beta}\, z'^2\,
 {if^{abc}\over N^2-1} S_{\mu\nu}(z') \times i
\int {d\lambda\over 2\pi} \int {d\mu\over 2\pi}
e^{-i\lambda/z}e^{-i\mu\left(1/z'-1/z\right)}\nn\\
&&\qquad\times 
\sum_X\la 0|
F_b^{(0)\,w\beta}(0)|hX\ra 
\la hX| gA_a^w(\lambda w)A_c^\alpha(\mu w)-
gA_a^\alpha (\lambda w) A_c^w(\mu w) |0\ra\nn\\
&&
+\int d\left({1\over z'}\right)\int d\left({1\over z}\right) \Omega^\mu_{\ \alpha}\Omega^\nu_{\ \beta} 
{z'^2 \over 1/z+i\epsilon}
{ if^{abc}\over N^2-1} S_{\mu\nu}(z') 
 \int {d\lambda\over 2\pi} \int {d\mu\over 2\pi}
e^{-i\lambda/z}e^{-i\mu\left(1/z'-1/z\right)}\nn\\
&&\qquad\times
\sum_X\la 0|
F_b^{(0)\,w\beta}(0)|hX\ra 
\la hX| gA_a^w(\lambda w)F_c^{(0)\,w\alpha}(\mu w)+
gF_a^{(0)\,w\alpha} (\lambda w) A_c^w(\mu w) |0\ra,
\eeq
where, in the first term, we have performed integration by parts for $\lambda$-integration,
which kills the factor ${1\over 1/z+i\epsilon}$.  
Integration over $1/z$ of this equation can be done immediately.  
The second term in $\left\{\quad\right\}$ of the last term of (\ref{SG5}) can be integrated parallelly.  
Following this procedure, $W_g^{\rm (b)}$ is obtained
by the integral of (\ref{SG5}) as
\beq
&&\hspace{-0.5cm}
W_g^{\rm (b)}={1\over 2} \int{d^4k\over (2\pi)^4}\int{d^4k'\over (2\pi)^4}
\left[ 
S_{L\,\mu\nu\lambda}^{abc}(k,k')
\widehat{\Gamma}_{L\, abc}^{\mu\nu\lambda}(k,k')\right]^{\rm twist-3}\nn\\
&&
=-\Omega^\mu_{\ \alpha}\Omega^\nu_{\ \beta}
\int d\left({1\over z'}\right) f^{abc} z'^2 S_{\mu\nu}\left({z'}\right)\nn\\
&&\qquad\qquad\times {1\over N^2-1} \sum_X\int{d\lambda\over 2\pi}e^{-i\lambda/z'}
\la 0|F_b^{(0)\,w\beta}(0)| hX\ra\la hX|gA_a^w(\lambda w) A_c^\alpha(\lambda w)|0\ra\nn\\
&&+\Omega^\mu_{\ \alpha}\Omega^\nu_{\ \beta}
\int d\left({1\over z'}\right) f^{abc} z'^2 S_{\mu\nu}\left({ z'}\right)\nn\\
&&\qquad\qquad\qquad\times {1\over N^2-1} 
\sum_X\int{d\lambda\over 2\pi}e^{-i\lambda/z'}
\la 0|F_b^{(0)\,w\beta}(0)| hX\ra\la hX|gF_a^{(0)\,w\alpha}(\lambda w) 
\int_\infty^\lambda d\mu A_c^w(\mu w)|0\ra\nn\\
&&+\Omega^\mu_{\ \alpha}\Omega^\nu_{\ \beta}\Omega^\lambda_{\ \gamma}
\int d\left({1\over z'}\right)z'^2 i f^{abc} \left. {\partial  S_{\mu\nu}\left(k'\right)
\over \partial k'^\lambda} \right|_{\rm c.l.}
\nn\\
&&\qquad\qquad\qquad\times  {1\over N^2-1} 
\sum_X\int{d\lambda\over 2\pi}e^{-i\lambda/z'}
\la 0|F_b^{(0)\,w\beta}(0)| hX\ra\la hX|\partial^\gamma
\left\{gA_a^w(\lambda w) A_c^\alpha(\lambda w)\right\}|0\ra\nn\\
&&-\Omega^\mu_{\ \alpha}\Omega^\nu_{\ \beta}\Omega^\lambda_{\ \gamma}
\int d\left({1\over z'}\right)z'^2 i f^{abc} \left. {\partial  S_{\mu\nu}\left(k'\right)
\over \partial k'^\lambda} \right|_{\rm c.l.}  {1\over N^2-1} 
\sum_X\int{d\lambda\over 2\pi}e^{-i\lambda/z'}
\la 0|F_b^{(0)\,w\beta}(0)| hX\ra\nn\\
&&\qquad\qquad\qquad\times
\la hX|\int_\infty^\lambda d\mu \left\{ 
\left(\partial^\gamma F_a^{(0)\,w\alpha}(\lambda w) \right) gA_c^w(\mu w)
+F_a^{(0)\,w\alpha}(\lambda w) g \partial^\gamma A_c^w(\mu w)
\right\}
|0\ra\nn\\
&&-{i\over 2}\,\Omega^\mu_{\ \alpha}\Omega^\nu_{\ \beta}\Omega^\lambda_{\ \gamma}
\int_1^\infty d\left({1\over z'}\right)\int_0^{1/z'} d\left({1\over z}\right)
{zz'\over 1/z'-1/z}
S_{L\,\mu\nu\lambda}^{abc}\left({1\over z},{1\over z'}\right)
\widehat{\Gamma}_{LF\,abc}^{\alpha\beta\gamma}\left({1\over z},{1\over z'}\right),
\label{SG6}
\eeq
where
\beq
&&\widehat{\Gamma}_{LF\,abc}^{\alpha\beta\gamma}\left({1\over z},{1\over z'}\right)\nn\\
&&\quad=\sum_X \int{d\lambda\over 2\pi}\int{d\mu\over 2\pi}
e^{-i\lambda/z}e^{-i\mu(1/z'-1/z)}
\la 0 |F^{(0)\,w\beta}_b(0)|hX\ra \la hX | F^{(0)\,w\alpha}_a(\lambda w) g 
 F^{(0)\,w\gamma}_c(\mu w)|0\ra.\qquad\ 
 \label{3Fcorr}
\eeq
The first four terms in (\ref{SG6}) come from the
combination of the first, third and the last terms in (\ref{SG5}),
while the last term in (\ref{SG6}) is from the second term of (\ref{SG5}). 
$\widehat{\Gamma}_{LF\,abc}^{\alpha\beta\gamma}\left({1\over z},{1\over z'}\right)$
can be identified as the lowest order part of the dynamical FF (\ref{gFraD}).  

Calculation of $W_g^{\rm (c)}$ can be performed in the same way.  The result reads
\beq
&&\hspace{-0.5cm}
W_g^{\rm (c)}={1\over 2}\int\frac{d^4k}{(2\pi)^4}\int\frac{d^4k'}{(2\pi)^4}
\biggl[ S^{abc}_{R\,\mu\nu\lambda}(k,k')\widehat{\Gamma}^{\mu\nu\lambda}_{R\,abc}(k,k')
\biggr]^{\rm twist-3}\nn\\
&&
=\Omega^\mu_{\ \alpha}\Omega^\nu_{\ \beta}
\int d\left({1\over z}\right) f^{abc} z^2 S_{\mu\nu}\left({z}\right)\nn\\
&&\qquad\qquad\times  {1\over N^2-1} 
\sum_X
\int{d\lambda\over 2\pi}e^{-i\lambda/z}
\la 0|A_b^{w}(0) g A_c^\beta (0)| hX\ra\la hX|F_a^{(0)\,w\alpha}(\lambda w) |0\ra\nn\\
&&+\Omega^\mu_{\ \alpha}\Omega^\nu_{\ \beta}
\sum_X
\int d\left({1\over z}\right) f^{abc} z^2 S_{\mu\nu}\left({ z}\right)\nn\\
&&\qquad\times {1\over N^2-1} \sum_X
\int{d\lambda\over 2\pi}e^{-i\lambda/z}
\la 0|\int^\infty_0 d\mu A_c^w(\mu w)F_b^{(0)\,w\beta}(0)| hX\ra\la hX|gF_a^{(0)\,w\alpha }(\lambda w) 
|0\ra\nn\\
&&-\Omega^\mu_{\ \alpha}\Omega^\nu_{\ \beta}\Omega^\lambda_{\ \gamma}
\sum
\int d\left({1\over z}\right)z^2 i f^{abc} \left. {\partial  S_{\mu\nu}\left(k\right)
\over \partial k^\lambda} \right|_{\rm c.l.}
\nn\\
&&\qquad\qquad\qquad\times {1\over N^2-1} \sum_X
\int{d\lambda\over 2\pi}e^{-i\lambda/z}
\la 0|A_b^{w}(0) gA_c^\beta (0) | hX\ra\la hX|\partial^\gamma F_a^{(0)\,w\alpha}(\lambda w) |0\ra\nn\\
&&-\Omega^\mu_{\ \alpha}\Omega^\nu_{\ \beta}\Omega^\lambda_{\ \gamma}
\int d\left({1\over z}\right)z^2 i f^{abc} \left. {\partial  S_{\mu\nu}\left(k\right)
\over \partial k^\lambda} \right|_{\rm c.l.}\nn\\
&&\qquad \times  {1\over N^2-1} \sum_X\int{d\lambda\over 2\pi}e^{-i\lambda/z}
\la 0| \int^\infty_0 d\mu gA_c^w(\mu w) F_b^{(0)\,w\beta}(0)| hX\ra
\la hX|  \partial^\gamma F_a^{(0)\,w\alpha}(\lambda w) 
|0\ra\nn\\
&&-{i\over 2}\,\Omega^\mu_{\ \alpha}\Omega^\nu_{\ \beta}\Omega^\lambda_{\ \gamma}
\int_1^\infty d\left({1\over z}\right)\int_0^{1/z} d\left({1\over z'}\right)
{zz'\over 1/z'-1/z}
S_{R\,\mu\nu\lambda}^{abc}\left({1\over z},{1\over z'}\right)
\widehat{\Gamma}_{RF\,abc}^{\alpha\beta\gamma}\left({1\over z},{1\over z'}\right),
\label{SG7}
\eeq
where
\beq
&&\widehat{\Gamma}^{\alpha\beta\gamma}_{RF\,abc}\left({1\over z},{1\over z'}\right)\nn\\
&&=\sum_X\int {d\lambda\over 2\pi}\int {d\mu\over 2\pi}e^{-i\lambda/z}e^{-i\mu (1/z'-1/z)}
\la 0| F_b^{(0)\,w\beta}(0) g F_c^{(0)\,w\gamma}(\mu w) |hX\ra \la hX|F_a^{(0)\,w\alpha}(\lambda w)|0\ra.\qquad
\label{3FRcorr}
\eeq
We can now compare the sum of $W_g^{\rm (b)}$ (\ref{SG6}) and $W_g^{\rm (c)}$ (\ref{SG7})
with (\ref{PPgfragForma}).  
We first remind that
the gluon's field strength is
$F_{a}^{\mu\nu}=\partial^\mu A_a^\nu - \partial^\nu A_a^\mu
+g f^{abc} A^\mu_b A^\nu_c=F^{(0)\mu\nu}_a+ gf^{abc}A_b^\mu A_c^\nu$.  
The first terms of (\ref{SG6}) and (\ref{SG7}) are the $O(g)$-contribution to (\ref{PPgfragForma})
from the $O(A^2)$-term of the field strength in the intrinsic FF (\ref{gFraI}).  
The second terms of  (\ref{SG6}) and (\ref{SG7}) are the $O(g)$-contribution
to (\ref{PPgfragForma})
from the gauge links in the intrinsic FF.  
The third terms of (\ref{SG6}) and (\ref{SG7}) are the $O(g)$-contribution to (\ref{PPgfragForma})
from the $O(A^2)$-term of the field strength in the kinematical FF (\ref{gFraK}).  
The fourth terms of  (\ref{SG6}) and (\ref{SG7}) are the $O(g)$-contribution
to (\ref{PPgfragForma})
from the gauge links in the kinematical FF.  
To identify the fifth terms of  (\ref{SG6}) and (\ref{SG7}), we note
the following relations:
\beq
&&\widehat{\Gamma}_{RF\,abc}^{\alpha\beta\gamma}\left({1\over z},{1\over z'}\right)
=\widehat{\Gamma}_{LF\,bac}^{\beta\alpha\gamma}\left({1\over z'},{1\over z}\right)^\star,\\
&&S^{abc}_{R\,\alpha\beta\gamma}\left({1\over z},{1\over z'}\right)
=S^{bac}_{L\,\beta\alpha\gamma}\left({1\over z'},{1\over z}\right)^\star.
\eeq
From these relations, the sum of the last terms in (\ref{SG6}) and (\ref{SG7})
are the $O(g)$ contribution to (\ref{PPgfragForma}) from the dynamical FF (\ref{gFraD}). 
This way, the basic formula  (\ref{PPgfragForma}) has been proved in 
the leading order with respect to the
QCD coupling constant. 

The method used here for the twist-3 gluon FF contribution to
$pp\to\Lambda^\uparrow X$ can also be applied to the twist-3 gluon distribution
function contribution to the double-spin asymmetry in
$\vec{p}p^\uparrow \to D X$, which occurs as a nonpole
contribution\cite{Hatta:2013wsa}.   Our method provides a clearer understanding for the
absence of the ghost-like terms in the corresponding LO twist-3 cross section.

\section{Summary}

In this paper we presented a new derivation of the basic formula (\ref{PPgfragForma}) for the twist-3 
gluon FF contribution to $pp\to \Lambda^\uparrow X$.
Our method uses the Ward-Takahashi identities for the partonic hard parts
from the outset
before performing the collinear expansion.   This method provides a robust shortcut to convert
the correlation functions of the gauge (gluon) fields
into the gauge invariant correlation functions for the gluon's field strengths.  
Furthermore it provides a clear understanding that the ghost-like terms appearing in the
Ward-Takahashi identities do not contribute to the LO twist-3 cross section.
Since this method is quite general, it will become a useful tool 
to extend the formula in the next-to-leading order calculation.

\section*{Acknowledgments}

This work has been supported by 
the establishment of Niigata university fellowships towards the creation of science
technology innovation (R.I.), the Grant-in-Aid for
Scientific Research from the Japanese Society of Promotion of Science
under Contract Nos.~19K03843 (Y.K.) and 18J11148 (K.Y.),
National Natural Science Foundation in China 
under grant No. 11950410495, Guangdong Natural Science Foundation under
No. 2020A1515010794
and research startup funding at South China
Normal University (S.Y.).

\section*{Appendix: Absence of the ghost-like contribution at LO twist-3}

Here we show that the ghost-like term (\ref{ghostfinal}) does not 
contribute to the twist-3 cross section.  
Actual forms of $G_{\mu\nu}^{abc}(k,k')$ are given in
(A7) for the $q\bar{q}\to gg$ channel, (A9) for the $qg\to gq$ channel, 
and (A10)-(A14) for the $gg\to gg$ channel in
\cite{Koike:2021awj}.  In all channels they take the structure
\beq
G_{\mu\nu}^{abc}(k,k')=(k^2g_{\mu\rho}-k_\mu k_\rho )f^{abc}\widetilde{G}^\rho_{\ \nu} (k,k'),
\eeq
where $\widetilde{G}^\rho_{\ \nu} (k,k')$ is some function of $k$ and $k'$ (and $xp$ and $x'p'$).  
Inserting this form into (\ref{ghostfinal}), one obtains
\beq
&&\hspace{-0.5cm}
\left[S_{L\,\mu\nu\lambda}^{abc}(k,k')
\widehat{\Gamma}_{L\, abc}^{\mu\nu\lambda}(k,k')\right]^{\rm ghost}\nn\\
&&=2\Omega^\mu_{\ \alpha}\Omega^\nu_{\ \beta}
{f^{abc}z'\over 1/z'-1/z} \left\{
-k_\mu \widetilde{G}_{P_h\nu}\left( {1\over z},{1\over z'}\right)
+2P_h\cdot k\widetilde{G}_{\mu\nu}\left( {1\over z},{1\over z'}\right)
-\Omega^\lambda_{\ \gamma}k^\gamma P_{h\mu}
\widetilde{G}_{\lambda\nu}\left( {1\over z},{1\over z'}\right)
\right.\nn\\
&&\left.\qquad\qquad
- {P_{h\mu}\over z}\left(
\left. \Omega^\lambda_{\ \gamma}k^\gamma
{\partial \widetilde{G}_{P_h\nu}(k,k')\over \partial k^\lambda}\right|_{\rm c.l.}
+ \left. \Omega^\lambda_{\ \gamma}k'^\gamma{\partial \widetilde{G}_{P_h\nu}(k,k')\over 
\partial k'^\lambda}\right|_{\rm c.l.}\right)
\right\}
\widehat{\Gamma}_{LFA\,abc}^{\alpha\beta w}(k,k').\qquad
\label{ghostfinal2}
\eeq
From this form one sees that all terms 
in $\left\{\cdots \right\}$ except for the first one
contribute only at twist-4:  
$\Omega^\mu_{\ \alpha}P_{h\mu}$ extracts ``$\alpha=-$" component from 
$\widehat{\Gamma}_{LFA\,abc}^{\alpha\beta w}(k,k')$
which is subleading, and 
$\Omega^\lambda_{\ \gamma}k^\gamma$ and $P_h\cdot k$, 
respectively, causes additional one- and two- power suppressions.  
For the first term which contributes at twist-3, 
we obtain from (\ref{FLderiv})
\beq
&&\Omega^\mu_{\ \alpha}k_\mu
\widehat{\Gamma}_{LFA\,abc}^{\alpha\beta w}(k,k')\nn\\
&&\quad=\sum_X\int d^4\xi\int d^4\eta e^{-ik\xi}e^{-i(k'-k)\eta}
\la 0| F^{(0)\,\beta w}_b(0)|hX\ra \la hX| 
(-i)
{\partial \over \partial \xi^\alpha}
F^{(0)\,\alpha w}_a(\xi) gA^w_c (\eta) |0\ra.  \qquad
\eeq
Here we note that the QCD equation of motion
$D_\alpha F^{\alpha w}_a+g\bar{\psi}t^a\wslash \psi=0$
implies
$\partial_\alpha F^{(0)\,\alpha w}_a(\xi)$ is of $O(g)$, and hence 
the first term in $\left\{\cdots \right\}$ of (\ref{ghostfinal2})
becomes $O(g^6)$.  
Consistent treatment of this term requires the inclusion of all $O(g^6)$ diagrams,
which is beyond the scope of this work.
We thus conclude that it does not contribute to the LO cross section.  
This proves that (\ref{ghostfinal}) does not contribute to the LO twist-3 cross section.



\begin{thebibliography}{99}

\bibitem{Koike:2021awj}
Y.~Koike, K.~Yabe and S.~Yoshida,
Phys. Rev. D \textbf{104}, no.5, 054023 (2021)
doi:10.1103/PhysRevD.104.054023
[arXiv:2107.03113 [hep-ph]].


\bibitem{Kanazawa:2001a} 
Y.~Kanazawa and Y.~Koike,
Phys.\ Rev.\ D {\bf 64}, 034019 (2001)

\bibitem{Zhou:2008} 
J.~Zhou, F.~Yuan and Z.~T.~Liang,
Phys.\ Rev.\ D {\bf 78}, 114008 (2008)

\bibitem{Koike:2015zya} 
Y.~Koike, K.~Yabe and S.~Yoshida,
Phys.\ Rev.\ D {\bf 92}, 094011 (2015)

\bibitem{Koike:2017fxr} 
  Y.~Koike, A.~Metz, D.~Pitonyak, K.~Yabe and S.~Yoshida,
  Phys.\ Rev.\ D {\bf 95}, no. 11, 114013 (2017)
  doi:10.1103/PhysRevD.95.114013
  [arXiv:1703.09399 [hep-ph]].

\bibitem{Gamberg:2018fwy}
L.~Gamberg, Z.~B.~Kang, D.~Pitonyak, M.~Schlegel and S.~Yoshida,
JHEP \textbf{01}, 111 (2019)
doi:10.1007/JHEP01(2019)111
[arXiv:1810.08645 [hep-ph]].

\bibitem{Koike:2022ddx}
Y.~Koike, K.~Takada, S.~Usui, K.~Yabe and S.~Yoshida,
Phys. Rev. D \textbf{105}, no.5, 056021 (2022)
doi:10.1103/PhysRevD.105.056021
[arXiv:2202.00338 [hep-ph]].

\bibitem{Ikarashi:2022yzg}
R.~Ikarashi, Y.~Koike, K.~Yabe and S.~Yoshida,
Phys. Rev. D \textbf{105}, no.9, 094027 (2022)
doi:10.1103/PhysRevD.105.094027
[arXiv:2203.08431 [hep-ph]].

\bibitem{Beppu:2010qn}
H.~Beppu, Y.~Koike, K.~Tanaka and S.~Yoshida,
Phys. Rev. D \textbf{82}, 054005 (2010)
doi:10.1103/PhysRevD.82.054005
[arXiv:1007.2034 [hep-ph]].
  
\bibitem{Mulders:2000sh} 
  P.~J.~Mulders and J.~Rodrigues,
  Phys.\ Rev.\ D {\bf 63}, 094021 (2001)
  doi:10.1103/PhysRevD.63.094021
  [hep-ph/0009343].

\bibitem{Koike:2019zxc}
Y.~Koike, K.~Yabe and S.~Yoshida,
Phys. Rev. D \textbf{101}, no.5, 054017 (2020)

\bibitem{Hatta:2013wsa} 
Y.~Hatta, K.~Kanazawa and S.~Yoshida,
Phys.\ Rev.\ D {\bf 88}, 014037 (2013)


\bibitem{Kanazawa:2013uia} 
K.~Kanazawa and Y.~Koike,
Phys.\ Rev.\ D {\bf 88}, 074022 (2013)


\end{thebibliography}
\end{document}